\documentclass{JHEP3}
\usepackage{amsmath}
\usepackage{axodraw}
\usepackage{epsfig}

\newcommand{\be}{\begin{equation}}
  \newcommand{\ee}{\end{equation}}
\newcommand{\bea}{\begin{eqnarray}}
  \newcommand{\eea}{\end{eqnarray}}
\newcommand{\nn}{\nonumber \\}

\newcommand{\ep}{\varepsilon}

\newcommand{\order}[1]{{\mathcal O}\left(#1\right)}
\newcommand{\scs}{\scriptscriptstyle}

\newcommand{\Leff}{{\mathcal L}_{\rm eff}}
\newcommand{\LQCDQED}{{\mathcal L}_{{\rm QCD} \times {\rm QED}} 
  (u, d, s,c, b)}
\def\dd{\mbox{d}}

\title
{\bf NNLO fermionic corrections to the charm quark
  mass dependent matrix elements in 
  {\boldmath $ \bar B \to X_s \gamma$}}

\author{Radja Boughezal$^{1}$, 
  Micha{\l} Czakon$^{1,2}$~and~Thomas Schutzmeier$^{1}$
  \\ \\
  $^1$~{Institut f\"ur Theoretische Physik
    und Astrophysik, Universit\"at W\"urzburg, \\
    Am Hubland, D-97074 W\"urzburg, Germany}
  \\ \\
  $^2$~{Department of Field Theory and Particle Physics,
    Institute of Physics, \\
    University of Silesia, Uniwersytecka 4, PL-40007 Katowice,
    Poland} }

\abstract{
  We compute the  virtual $\order{\alpha_s^2}$ fermionic contributions to
  the charm quark mass dependent matrix elements  of the $ \bar B \to X_s
  \gamma$ decay. In the case of a massless quark loop insertion into the gluon
  propagator, our result obtained  as an expansion in $z=m_c^2/m_b^2$ and  an
  exact expression in terms of multi-fold MB integrals, confirms the findings
  of Bieri, Greub and Steinhauser ~\cite{Bieri:2003ue}. We observe, however,
  large deviations in the case of a b-quark loop insertion. The charm quark
  loop shows smaller, but still noticeable differences.
}

\preprint{ arXiv:0707.3090 [hep-ph] }

\begin{document}

\section{Introduction}
\label{sec:intro}

One of the most interesting rare B-meson decays is the inclusive 
$\bar B\, \to X_s \gamma $ mode which provides precise and
clean short-distance information on 
$\Delta B \,=\, 1$ flavor changing neutral currents (FCNCs). 
This process occurs in the SM only at the loop level,
through the exchange of W-bosons and up-type quarks, making
it highly sensitive to non-standard effects which are not suppressed
by additional factors $\alpha$ relative to the SM
contributions. Combined with
the low sensitivity of $B \to X_s \gamma$ to non-perturbative
effects, this makes it possible to observe new physics contributions
indirectly, or to set limits on the relevant masses and
coupling parameters. In fact 
the decay width~$ \Gamma\left( \bar B \to X_s \gamma\right)$ 
is well approximated by the partonic decay 
rate $\Gamma( b \to X_s^{parton} \gamma) $ 
which can be analyzed within the framework of renormalization-group-improved
perturbation theory\,.

\medskip

In view of the above, it is obvious
that both accurate measurements and precise theoretical SM calculations,
with a good control of both perturbative and non-perturbative corrections,
have to be provided.
As far as the experimental side is concerned, measurements are performed by
several B physics experiments operating within various experimental settings: 
CLEO~\cite{ref_cleo} (Cornell), BaBaR~\cite{ref_babar} (SLAC), 
Belle~\cite{ref_belle} (KEK) and ALEPH~\cite{ref_aleph} (CERN)\,.
Combining the measurements of the first three (the last one has very large
error bars) for the branching ratio 
${\mathcal B}(\bar B\to X_s\gamma)$
leads to a world average with a cut $E_{\gamma} > 1.6\;$GeV 
in the ${ \bar B}$-meson rest frame which reads~\cite{Group:2007cr}
\be \label{eq:HFAG}
{\mathcal B}(\bar B \to X_s \gamma)_{\scs E_{\gamma} > 1.6\,{\rm GeV}}^{\scs\rm exp
}
= \left(3.55\pm 0.24{\;}^{+0.09}_{-0.10}\pm0.03\right)\times 10^{-4},
\ee
where the first error is given by the statistic and systematic uncertainty,
the second one is due to the theory input on the shape function, and the third
one is caused by the $b \to d\gamma$ contamination\,. This average
is in good agreement with the recent theoretical 
estimate including known next-to-next-to-leading-order (NNLO) effects
~\cite{Misiak:2006zs}
\be 
\label{theoretical B}
{\mathcal B}({\bar B}\to X_s\gamma)_{\scs E_{\gamma} > 1.6\,{\rm GeV}}^{\scs\rm theo} = (3.15 \pm 0.23) \times 10^{-4},
\ee
where the error consists of four types
of uncertainties added in quadrature: non-perturbative (5\%),
parametric (3\%), higher-order (3\%) and $m_c$-interpolation ambiguity (3\%).

The total experimental error of about 7\% in Eq.~(\ref{eq:HFAG}) 
is of the same size as the expected ${\mathcal O}(\alpha_s^2)$
corrections to the perturbative transition 
$b \to X_s^{\rm parton}\gamma$~\cite{Kagan:1998ym}, which calls for
completing the SM calculations with this accuracy level. 

\medskip

QCD corrections to the partonic decay rate 
$\Gamma\left( b \to s \gamma\right) $ contain large logarithms
of the form $\alpha_s^n\left(m_b\right) \, \ln^m\left(m_b/M\right)$, where
$M\,=\,m_t$ or $M\,=\,m_W$ and $m\le n$ (with $n\,~=~\, 0, 1, 2, \dots$)\,.  
In order to get a reasonable prediction for the decay rate with
next-to-next-leading-log (NNLL) precision, it is necessary to resum
logarithms with ($m \,=\,n\,, n-1,\, n-2$) with the help of 
renormalization-group techniques.
A convenient framework 
is an effective low-energy theory obtained from the SM by decoupling
the heavy electroweak bosons and the top quark\,. The resulting effective 
Lagrangian, given in the next section, is a product of 
the Wilson coefficients $C_i(\mu)$ which play the role of coupling
constants, and local flavor-changing operators $Q_i(\mu)$\,.

\medskip

As far as the next-to-leading order corrections are concerned,  
the program has been completed already a few years ago, 
thanks to the joint effort of many groups 
(see e.g.~\cite{Buras:2002er,Hurth:2003vb} and references 
therein). Moreover, each of the ingredients has been cross-checked by more 
than one group\footnote{In fact the complete calculation of Bremsstrahlung
  corrections \cite{Pott:1995if} has not been checked, but the effects are
  extremely small.}. The next-to-next-to-leading order calculation, which involves 
hundreds of three-loop on-shell vertex-diagrams and thousands of four-loop
tadpole-diagrams, is a very complicated task and is currently 
under way\,. A consistent calculation of $ b \to s \gamma$ at this
order requires three steps:  
\begin{itemize}
\item Matching:~ Evaluation of $C_i(\mu_0)$ at the renormalization scale
  $\mu_0 \sim M_W,m_t$~ by requiring equality of the SM 
  and effective theory Green's functions at the leading order 
  in (external momenta)$/(M_W,m_t)$ to  $\order{\alpha_s^2}$\,.
  All the relevant Wilson coefficients have already been 
  calculated~\cite{Bobeth:1999mk,Misiak:2004ew} to this precision, 
  by matching the four-quark operators $Q_1,\dots,Q_6$ and the dipole operators 
  $Q_7$ and $Q_8$ at the two- and three-loop level
  respectively.
\item Mixing:~ Calculation of the operator mixing under renormalization, by
  deriving the effective theory Renormalization Group Equations (RGE) and
  evolving $C_i(\mu)$ from $\mu_0$ down to the low-energy 
  scale $\mu_b \sim m_b$, using the anomalous-dimension matrix (ADM)
  to $\order{\alpha_s^3}$\,.
  Here, the three-loop renormalization in the $\{Q_1, \dots ,Q_6\}$ and  
  $\{Q_7, Q_8 \}$ sectors was found in~\cite{Gorbahn:2004my,Gorbahn:2005sa}, 
  and results for the four-loop mixing 
  of $Q_1, \dots, Q_6$ into 
  $Q_7$ and $Q_8$ were recently provided in~\cite{Czakon:2006ss}, thus
  completing the anomalous-dimension matrix.
\item Matrix elements:~ Evaluation 
  of the on-shell $b \to s\gamma$ amplitudes 
  at~ $\mu_b~\sim~m_b$\, to $\order{\alpha_s^2}$. This task is not complete
  yet, although a number of contributions is known.
  The two-loop matrix element of the photonic
  dipole operator $Q_7$, together with the corresponding 
  bremsstrahlung, was found in~\cite{Melnikov:2005bx,Blokland:2005uk},
  confirmed in~\cite{Asatrian:2006ph} and subsequently extended to include the
  full 
  charm quark mass dependence in~\cite{Asatrian:2006rq}.
  In~\cite{Bieri:2003ue}, the $\order{\alpha_s^2\,n_f}$ contributions were
  found to the two-loop matrix elements of $Q_7$ and $Q_8$, as well as to the
  three-loop matrix elements of $Q_1$ and $Q_2$, using an expansion
  in the quark mass ratio $m_c^2/m_b^2$.
  Diagrammatically,
  these parts are generated by inserting a one-loop quark
  bubble into the gluon propagator of the two-loop Feynman diagrams.
  Naive non-abelianization (NNA) is then used to get an estimate
  of the complete corrections of $\order{\alpha_s^2}$ by replacing 
  $n_f$ with $-\frac{3}{2}\beta_0$. Moreover, the contributions of the
  dominant operators at $\order{\alpha_s^2 \beta_0}$ to the photon energy spectrum  
  have been computed in~\cite{Ligeti:1999ea}.
  Finally, in~\cite{Misiak:2006ab}, the full matrix elements of $Q_1$ and
  $Q_2$ have been computed in the large $m_c$ limit, $m_c\gg
  m_b/2$. Subsequently, an interpolation 
  in the charm quark mass has been done down to the physical region,
  under the assumption that the $\beta_0$-part is a good approximation at
  $m_c\,=\,0$\,. This is the source of the interpolation uncertainty
  that has been mentioned below Eq.~(\ref{theoretical B}). 
\end{itemize}
To date, no independent check has been provided for
the $\order{\alpha_s^2\,n_f}$ results for the $m_c$-dependent
matrix elements of $Q_1$ and $Q_2$, despite the fact that they constitute a major
input both for the NNA and for the interpolation of the non-NNA terms between
$m_c \gg m_b/2$ and $m_c < m_b/2$, and are thus crucial for the accuracy of
Eq.~(\ref{theoretical B}). Besides, for the numerical
evaluation, it was assumed that $n_f \,=\, 5$ massless fermions
are present in the quark loop. 
Since the charm and bottom quark masses are not negligible, 
it is important to check their numerical relevance.
\medskip

In this paper, we present a calculation of the virtual 
$\order{\alpha_s^2 n_f}$
contribution to the matrix elements of the operators $Q_1$ and $Q_2$.
As far as contributions from diagrams with a massless quark loop 
insertion into gluons are concerned, our result was obtained 
using two different methods: 
an expansion in the mass ratio $m_c^2/m_b^2$,
which confirms the findings of~\cite{Bieri:2003ue}, 
as well as an exact evaluation in terms of multi-fold Mellin Barnes (MB)
numerical integrals. The complexity of diagrams with a massive loop 
insertion (charm and bottom) required a mixed approach, where both  
MB techniques and differential equations were used numerically.  

\medskip 

This paper is organized as follows\,. In the next section, we introduce
the relevant effective Lagrangian and present the methods used. Subsequently, our results
for the matrix element of the operator $Q_2$ ($Q_1$ is related to $Q_2$
by a color factor at this level of perturbation theory) are given in the form 
of fitting formulae, which satisfactorily approximate both the massless and the massive
cases over the full experimentally interesting range of values of the
charm and bottom quark mass ratio.
Finally, we give our conclusions. An appendix contains the behavior of the
contributions in the limit $m_c \gg m_b$.
\section{Calculation}
\label{calculation}
As mentioned in the introduction, we work within 
an effective low-energy theory with five active quarks, obtained
from the SM by integrating out the heavy degrees of freedom with 
a mass scale $M \ge M_W$. The Lagrangian of such a theory reads
\be \label{eq:effectivelagrangian}
\Leff = \LQCDQED + \frac{4G_F}{\sqrt{2}} V^\ast_{ts} V_{tb} 
\sum^{8}_{i= 1} C_i (\mu) \, Q_i (\mu).
\ee
Here the first term is the usual QCD-QED Lagrangian for the light SM
quarks. In the second term, $V_{ij}$ denotes the elements 
of the Cabbibo-Kobayashi-Maskawa matrix, $G_F$ is the Fermi coupling 
constant and $C_i(\mu)$ are the Wilson coefficients 
of the corresponding operators $Q_i$ evaluated at the scale $\mu$.
Adopting the operator definitions of~\cite{Chetyrkin:1996vx}, 
the physical operators that are relevant for our calculation read

\bea \label{eq:physicaloperators}
Q_1 \, & = & \, (\bar{s}_L \gamma_\mu T^a c_L) (\bar{c}_L 
\gamma^\mu T^a b_L), \nn 
Q_2 \, & = & \, (\bar{s}_L \gamma_\mu c_L) (\bar{c}_L 
\gamma^\mu b_L) ,\nn 
Q_4 \, & = & \, (\bar{s}_L \gamma_\mu T^a b_L) \sum\nolimits_q (\bar{q}
\gamma^\mu T^a q) , \nn 
Q_7 \, & = & \, \frac{e}{g_s^2}\,{\overline m_b}(\mu)\, (\bar{s}_L 
\sigma^{\mu \nu} b_R) \, F_{\mu \nu} ,
\eea
where $Q_4$ and $Q_7$ are needed for the renormalization, and
$\overline{m_b}(\mu)$ is the b-quark $\overline{MS}$ mass. 
The sum over $q$ runs over all light quark fields, and
e and $g_s$ are the electromagnetic and strong coupling constants
respectively. $F_{\mu \nu}$  
is the electromagnetic field strength tensor, 
and $T^a$ ($a\,=\, 1 \dots 8$) denote the $SU(3)$ color generators.

\medskip

The choice of the basis of the local four-quark operators
is made such that no problems connected with the treatment of $\gamma_5$
in $d\,=\,4-2\epsilon$ arise~\cite{Chetyrkin:1996vx}. 
This allows for consistent use of a fully
anticommuting $\gamma_5$ in dimensional regularization throughout
the whole calculation.
The only evanescent operator needed for renormalization
would in principle be $Q_{11}$ defined as
\be
Q_{11}  = (\bar{s}_L \gamma_\mu \gamma_\nu \gamma_\rho T^a c_L)
(\bar{c}_L \gamma^\mu \gamma^\nu \gamma^\rho T^a b_L) - 16\, Q_1 ,
\ee
but it turns out that it does not contribute to the fermionic part.

\medskip 

Since one can show that at this order, the contributions
to the matrix elements of the operator $Q_1$ are related to those
of $Q_2$ by a simple substitution of color factors
\be
\langle s \gamma|Q_{1}|b \rangle  = -\frac{1}{2\, N_c}
\langle s \gamma|Q_{2}|b \rangle  
\ee     
we concentrate
in the following on the derivation of the result for $Q_2$.
\begin{figure}
  \begin{center}
    \epsfig{file=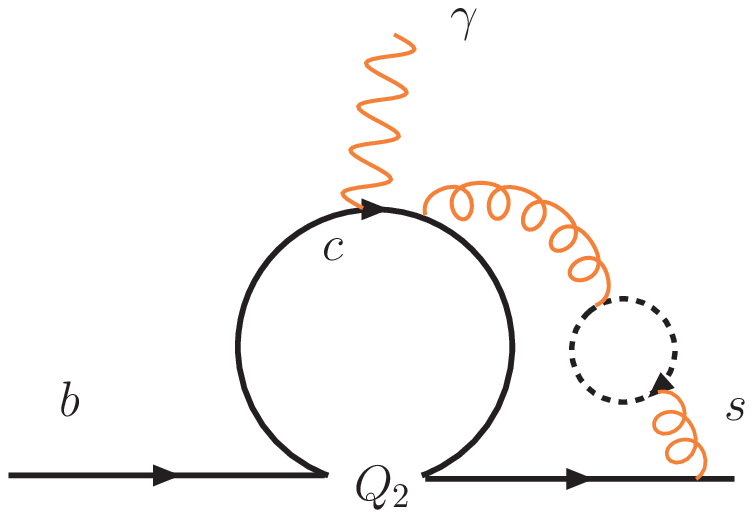, width=.35\textwidth}
    \epsfig{file=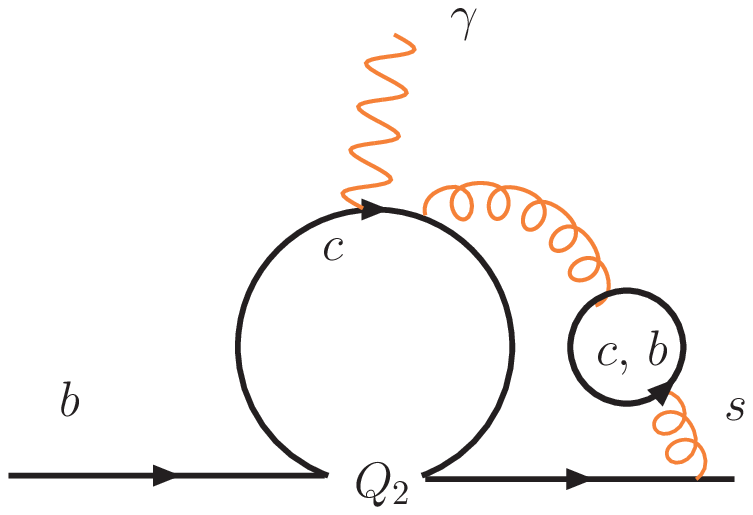, width=.35\textwidth}
    \caption{\label{sample Q2}\sf Typical $3$-loop vertex graphs calculated
      in this paper. The dotted bubble refers 
      to a massless quark. \sf}
  \end{center}
\end{figure}
Upon reducing the set of scalar integrals involved in the calculation 
of diagrams with a $Q_2$ insertion, of which a sample is shown in 
FIG.~\ref{sample Q2}, we are left with a set of $18$~master integrals
in the case with a massless quark loop insertion into the gluon 
propagator, $47$~master integrals in the case with a massive 
b-quark loop insertion and $38 $ for the c-quark case.
These master integrals have been calculated by combining 
two different approaches.

\medskip

In the first approach, we have used the Mellin-Barnes 
technique~\cite{Smirnov:1999gc,Tausk:1999vh}, which relies 
on the identity
\bea
\frac{1}{(X+Y)^{\lambda}} =
\int_{\beta - i  \infty}^{\beta + i  \infty}
\frac{Y^z}{X^{\lambda+z}} \frac{\Gamma(\lambda+z) \Gamma(-z)}
{\Gamma(\lambda)} \frac{\dd z}{2\pi  i} ,
\eea
where $-{\rm Re}\, \lambda <\beta <0$. With this relation  
a sum of terms raised to some power is replaced with a product 
of factors. We have derived our MB representations for the master integrals
using the automated package~\cite{chachamis-czakon:2006}, and used 
the {\tt MB} package of ref.~\cite{Czakon:2005rk} for their analytic 
continuation. In the case where the inserted quark loop into
the gluon propagator is massless, we have used two methods.
In the first, we have performed an expansion in the quark mass ratio 
$z= m_c^2/m_b^2$ by closing contours. 
The coefficients of this expansion were given
by (at most) one-dimensional MB integrals which have subsequently
been expressed as a sum over residues and resummed with the help of {\tt XSummer}~\cite{Moch:2005uc}.
Our result for the matrix element of $Q_2$ using this method
is in full agreement with~\cite{Bieri:2003ue}. It is also
consistent with the second method, where the exact
$z$-dependence was retained and a numerical integration of the MB
representations was performed using the {\tt MB} package.

\medskip

Due to poor convergence, it turned out not to be possible to compute 
all the master integrals 
that occur in diagrams with a massive 
quark loop insertion with the help of MB representations. 
For these particular cases, we have used a second approach based on the
method of differential  equations.
Using the fact that the master integrals $V_i(z,\epsilon)$ (after
rescaling by a trivial factor) are
functions of $\epsilon$ and the mass ratio $z$ or its inverse
$y=z^{-1}$, respectively, a system of differential equations
has been generated,
\begin{equation}
  \frac{d}{dy}V_i(y,\epsilon) = A_{ij}(y,\epsilon) V_j(y,\epsilon),
\end{equation}
where the right-hand side was again expressed through master integrals
with the help of relations obtained from the reduction. The
block-triangular matrix $A_{ij}(y,\epsilon)$ is composed of rational
functions of $\epsilon$ and $y$.

\medskip

The solution of this system for arbitrary values of $y$ proceeded
in two steps. First, an expansion in $\ep$ and $y$ for $\epsilon,\,
y\rightarrow 0$ was performed with the ansatz
\begin{eqnarray}
  V_i(y,\epsilon) &=& \sum_{nmk} c_{inmk} \epsilon^n y^m
  \log^ky, \quad
  n,m=-3,-2,\dots;\;\;k=0,\dots,3+n\,\Theta(n)
\end{eqnarray}
and the coefficients were calculated recursively up to high
powers of $y$~\cite{Boughezal:2006uu}. In this
limit, $m_c^2 \gg m_b^2$ and the initial conditions
correspond to vertex diagrams that can be derived in an
automated way from diagrammatic large-mass expansions. With
this series at hand we were able to compute the master
integrals for $y\ll 1$ with high precision.
In the second step a numerical integration of the
$\epsilon$-expanded system into the physical region, $y>1$, was
carried out using the high precision values at a starting point
$y \ll 1$. For this task, we have used the Fortran package ODEPACK
\cite{odepack}, which allows for numerical solutions of
huge systems of differential equations. To avoid points of
numerical instability on the real axis, the integration
path has been shifted into the complex plane.

\medskip

Besides the bare three-loop matrix elements, we also needed matrix elements of
two types of counterterms: those coming from the renormalization of the strong
coupling constant  $\alpha_s$ (known from the QCD beta function \cite{van
Ritbergen:1997va,Czakon:2004bu}), and those due to the mixing of $Q_2$ into
other operators.  For the corresponding details, we refer the reader
to~\cite{Bieri:2003ue}.  We just mention here that the renormalization
constants used for the renormalization of the diagrams with a massless quark
loop insertion into the gluon propagator are  the same as those used for the
renormalization of the diagrams with a massive quark loop insertion, because
all the needed renormalization constants were derived in the
$\overline{MS}$-scheme and are therefore  mass independent.
\section{Results}
\label{results}
\begin{figure}
  \centerline{\epsfig{file=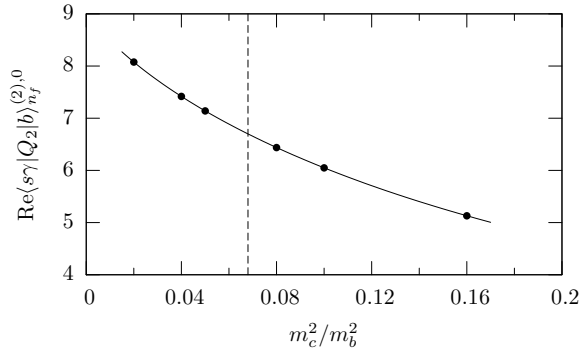, width=.5\textwidth}}
  \caption{  \label{masslessPlot}\sf Plot of $\mbox{Re}\langle 
    s\gamma|Q_2|b\rangle^{(2),0}_{n_f}$ for one massless flavor and
    $\mu_b=m_b$. The dashed
    vertical line corresponds to the central value $m_c^2/m_b^2=0.068$. \sf}
\end{figure}

We now present the results for the amplitudes of the three different 
quark loop insertions with the following normalization
\be
\langle s\gamma|Q_2|b\rangle_{{\mathcal O}(\alpha_s^2 n_f)} = 
\left(\frac{\alpha_s}{4\pi}\right)^2\,\frac{e}{8\pi^2}\,m_b\,\,n_f\,
\langle s\gamma|Q_2|b\rangle^{(2),M}_{n_f}\,\,
\bar{u}_s \, R\,\varepsilon\hspace{-0.45em}/ \,q\hspace{-0.45em}/\, u_b
\ee
where $m_b$ denotes the b-quark pole mass, 
$\varepsilon$ and $q$ are the photon
polarization and momentum, $R=(1+\gamma_5)/2$ is the right handed projection
operator, and $n_f$ is the number 
of active flavors of a given mass. The superscript~$(2)$ counts
the powers of $\alpha_s$ and 
$M=\,$($0,\, m_b\, \mbox{or}\, m_c$) denotes the mass
of the quark running in the loop inserted into the gluon propagator.      

\medskip

It is important that there is no need for a b-quark mass
renormalization in our calculation, so we use the different schemes as guided
by the complete calculation, and thus turn to the 1S mass \cite{Hoang:1999ye}
for our final study.

\medskip

\begin{figure}
  \begin{minipage}{1.\textwidth}
    \begin{center}
      \begin{minipage}{.49\textwidth}
        \epsfig{file=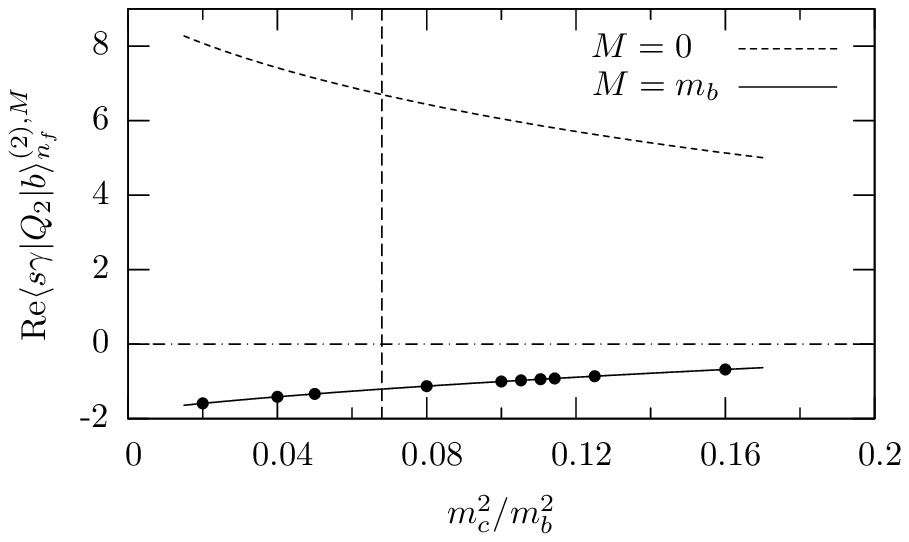, width=1.\textwidth}
        \centerline{(a)}
      \end{minipage}
      \begin{minipage}{.49\textwidth}
        \epsfig{file=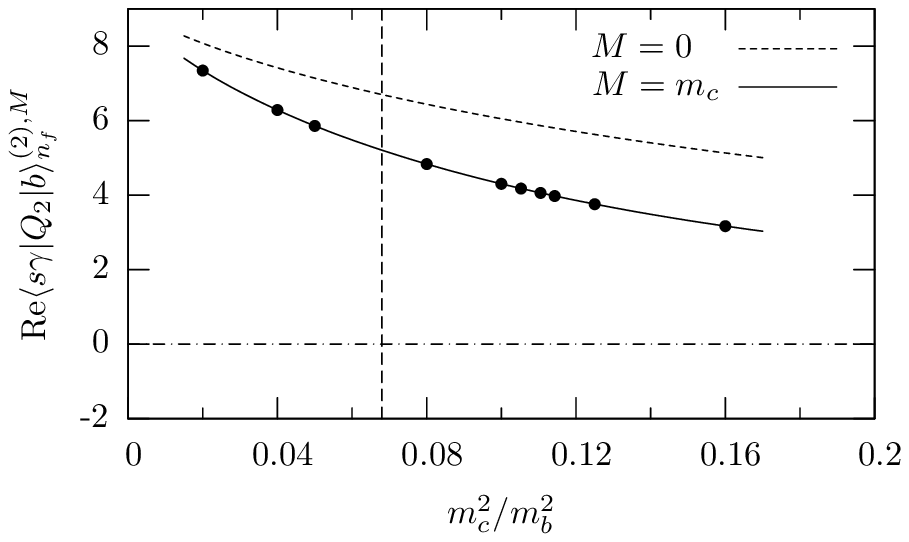, width=1.\textwidth}\hfill
        \centerline{(b)}
      \end{minipage}
    \end{center}
  \end{minipage}
  \caption{\label{Zdependence}\sf Plots of $\mbox{Re}\langle
    s\gamma|Q_2|b\rangle^{(2),M}_{n_f}$ as
    function of  $m_c^2/m_b^2$ with $M=m_b$ (a) and
    $M=m_c$ (b) and $\mu_b=m_b$. For comparison, we also show the $M=0$ case.\sf}
\end{figure}

All our subsequent results are given in the form of fitting formulae that cover the
whole interesting range of variation of $z$. With the current input given in
Table~\ref{tab:exp.input}, and allowing for a 3 sigma variation of $m_c(m_c)$
and $m_b^{1S}$, this corresponds to $z \in [0.017,0.155]$. In the plots of
Fig.~\ref{masslessPlot} and Fig.~\ref{Zdependence}, we also show the central
value, which is currently $z = 0.068$.
\begin{table}[t]
  \begin{center}
    \begin{tabular}{|l||l|}
      \hline
      Input parameter &  experimental value \\[1mm]\hline \hline
      $m_b^{1S}$
      & $(4.68 \pm 0.03)\;{\rm GeV}$~\cite{Bauer:2004ve} 
      \\[1mm]
      $m_c(m_c)$
      & $(1.224 \pm 0.017 \pm 0.054)\;{\rm GeV}$~\cite{Hoang:2005zw} 
      \\\hline
    \end{tabular}
  \end{center}
  \caption{\sf~Experimental inputs relevant for the present calculation
    \label{tab:exp.input} }
\end{table}

\medskip

In the case of a massless quark loop insertion we agree with the previous
calculation \cite{Bieri:2003ue}. Nevertheless we present here a fitting
formula for easier comparison with our new results
\begin{eqnarray}\label{masslessFit}
       \mbox{Re}\langle s\gamma|Q_2|b\rangle^{(2),0}_{n_f} &=& 
       9.080 - 0.7624\,z -5.069\,z^2+12.61\,z\,\ln z
       \nonumber\\
       &+& (-9.679 + 5.157\,z +1.726 \,z^2 - 16.18\,z\,\ln z )\ln(m_b/\mu) \nonumber\\
      &+& \frac{800}{243}\,\ln^2(m_b/\mu)
\end{eqnarray}
In Fig.~\ref{masslessPlot} we show the data points used to obtain the fit
function together with the result Eq.~\ref{masslessFit}.
Motivated by the presence of logarithms in the small $z$ expansion,
we have included terms proportional to $z\,\ln z$ above, and in
equations~(\ref{mbFitFormula}) and (\ref{mcFitFormula})
below, which improved the quality of the fit significantly.
As it stands, our fit function reproduces the exact values with a relative
precision of at least $10^{-4}$.

\medskip

Our result for the contribution of the diagrams with a massive 
b-quark loop insertion is given by the fitting formula of the same relative
precision as above
\begin{eqnarray}\label{mbFitFormula}
       \mbox{Re}\langle s\gamma|Q_2|b\rangle^{(2),m_b}_{n_f} &=& 
       -1.836 +2.608 \,z +0.8271\,z^2-2.441\,z\,\ln z
       \nonumber\\
       &+& (-9.595 + 5.157\,z +1.726 \,z^2 -16.18\,z\,\ln z) \ln(m_b/\mu) \nonumber\\
      &+& \frac{800}{243}\,\ln^2(m_b/\mu)
\end{eqnarray}
which is plotted in Fig.~\ref{Zdependence}(a). It is interesting
to note that the massless approximation overestimates
the massive result by a factor of $6$ and, moreover, has the opposite
sign. This points to an expected decoupling like effect. Remark, however, that
the scale dependence is just as strong as in the massless case.

\medskip

When the massless approximation is confronted with the result for the charm
quark loop insertion, one still observes noticeable differences as shown in
Fig.~\ref{Zdependence}(b). The obtained fitting formula of relative precision
of $10^{-4}$ is
\begin{eqnarray}\label{mcFitFormula}
       \mbox{Re}\langle s\gamma|Q_2|b\rangle^{(2),m_c}_{n_f} &=& 
       9.099 +13.20\,z -19.68 \,z^2+25.71\,z\,\ln z
       \nonumber\\
       &+& (-9.679 + 13.62\,z -13.94\,z^2 -12.98\,z\,\ln z )\ln(m_b/\mu) \nonumber\\
      &+& \frac{800}{243}\,\ln^2(m_b/\mu)
\end{eqnarray}
When $z$ approaches $0$ both Eq.~\ref{masslessFit} and
Eq.~\ref{mcFitFormula} tend approximately to the same value, as it should
be. The residual difference is due to the quality of the fit outside the validity
range, and the steepness of the curves in Fig.~\ref{Zdependence}.

\medskip

Let us finally comment on the renormalization group logarithms in
Eqs.~\ref{masslessFit} to \ref{mcFitFormula}. It has already been shown in the
massless case in \cite{Misiak:2006ab} that the single logs can be expressed
through the $a(z)$ and $b(z)$ functions known from the NLO calculation
\cite{Buras:2002tp}. In fact the exact expression for the coefficient of
$\log(m_b/\mu)$ in this case is
\begin{equation}
\frac{8}{3} \left( \Re\left( a(z) + b(z)\right) - \frac{290}{81} \right).
\end{equation}
Using the RGE one can show that the coefficient of the same log in
Eq.~\ref{mbFitFormula} is\footnote{We thank M. Misiak for pointing this out.}
\begin{equation}
 \frac{8}{3} \left( \Re\left( a(z) + b(z) + b(1)\right) - \frac{290}{81} \right),
\end{equation}
and similarly for Eq.~\ref{mcFitFormula}
\begin{equation}
 \frac{8}{3} \left( \Re\left(a(z) + 2 b(z)\right) - \frac{290}{81} \right).
\end{equation}

\section{Conclusions}

In this paper, we have presented a calculation 
of the ${\mathcal O}(\alpha_s^2 n_f)$ contribution
to the charm quark mass dependent matrix elements 
of the $ \bar B \to X_s \gamma$ decay. In~\cite{Bieri:2003ue}, a similar 
calculation was done as a mass expansion in the ratio $z=m_c^2/m_b^2$
assuming that the b and c quarks inside the fermionic loop inserted
into the gluon propagator are massless. The results have
then been used to estimate the complete corrections 
of ${\mathcal O}(\alpha_s^2)$ to the matrix elements
$\langle s \gamma|Q_{1,2}|b \rangle$ by replacing $n_f$ with 
$-\frac{3}{2}\beta_0$, i.e. following the naive non-abelianization procedure.
They form an important part of the NNLO contributions used in the recent 
theoretical estimate of the branching ratio 
${\mathcal B}(\bar B\to X_s\gamma)$~\cite{Misiak:2006zs}.     
Our goals were firstly to provide an independent check of the 
${\mathcal O}(\alpha_s^2 n_f)$ corrections for massless quarks,
and secondly to check the validity of the massless approximation
by keeping the full mass dependence. These goals have been reached, and
our result in the massless approximation, which we have obtained
as a mass expansion in $z$ as well as an exact evaluation in terms 
of multi-fold MB integrals, confirms the findings of~\cite{Bieri:2003ue}.
However, although the massless approximation reproduces the contribution of
the diagrams with a massive c-quark loop insertion 
reasonably well, it is not justified for the diagrams with a massive
b-quark. Neglecting the mass
of this particle leads to an estimate of its contribution that is 
$6$ times larger than its true value and moreover has the opposite 
sign. Of course, since this effect comes from a single quark family its impact
on the branching ratio is rather mild. It turns out that, when
compared with the interpolation from \cite{Misiak:2006ab}, one finds an
enhancement between one and two percent, depending on the 
renormalization scale $\mu$.

\section*{Acknowledgments}

We would like to thank M. Misiak for useful comments, which helped to
substantially improve the manuscript and A. Hoang for a discussion on the
b-quark mass definition.

This work was supported by the
Sofja Kovalevskaja Award of the Alexander von Humboldt Foundation
sponsored by the German Federal Ministry of Education and Research.

\appendix

\section{Contributions in the limit $m_c \gg  m_b$}

During our calculation we also derived the large $m_c$ limit of all the
contributions. In the massless case, we agree with \cite{Misiak:2006ab}. For
the two massive cases, we have (with $\mu=m_b$)
    \begin{eqnarray}
     \label{LargeMc:CaseMb}
       \langle s\gamma|Q_2|b\rangle^{(2),m_b}_{n_f} &=&
       4.25648 + 0.503085\,\ln z + 0.888889\, \ln^2 z
       \nonumber\\
       &+& \frac{1}{z}\,(-0.725053 - 1.80916\,\ln z + 0.0938272\,\ln^2 z )
       \nonumber\\
       &+& \frac{1}{z^2}\,(-1.39486 - 0.968501\,\ln z - 0.147443\,\ln^2 z)
        +  {\mathcal O}\left(\frac{1}{z^3}\right), \nonumber \\
    \end{eqnarray}
and
    \begin{eqnarray}
     \label{LargeMc:CaseMc}
       \langle s\gamma|Q_2|b\rangle^{(2),m_c}_{n_f} &=&
       1.67932 + 0.526749\,\ln z + 0.823045\, \ln^2 z
       \nonumber\\
       &+& \frac{1}{z}\,(0.20839 + 0.11775\,\ln z + 0.128395\,\ln^2 z )
       \nonumber\\
       &+& \frac{1}{z^2}\,(-0.0360638 - 0.0470166\,\ln z + 0.0324515\,\ln^2 z)
        +  {\mathcal O}\left(\frac{1}{z^3}\right). \nonumber \\
    \end{eqnarray}

\end{document}